\documentstyle[12pt]{article}

\begin{document}

\pagenumbering{roman} 

 

\begin{center}
{\em Quantum anharmonic oscillator and its statistical properties}
\end{center}

\begin{center}
Maciej M. Duras
\end{center}

\begin{center}
Institute of Physics, Cracow University of Technology, 
ulica Podchor\c{a}\.zych 1, PL-30084 Cracow, Poland.
\end{center}

\begin{center}
Email: {\tt mduras @ riad.usk.pk.edu.pl}
\end{center}

\begin{center}
AD 2007 November 21st
\end{center}

\begin{center}
\end{center}

\begin{abstract}
In the present article a family of quantum anharmonic oscillators is studied 
using Hermite's function basis (Fock's basis) in the Hilbert space. 
The numerical investigation of the eigenenergies of that family is presented. 
The statistical properties of the calculated eigenvalues are compared 
with the theoretical predictions derived from the Random Matrix Theory. 
Conclusions are inferred. 
\end{abstract}

\section{Motivation of the present work}

The simple quantum harmonic oscillator is an excellent model of many physical systems,
{\sl e.g.} electromagnetic field. The excitation particles or quanta of the  
electromagnetic field are called photons \cite{Einstein1905,Lewis1926}
Moreover it applies for systems of atoms (ions, nuclei) in ideal crystals
interacting via harmonic attractive force, \, {\sl etc.}
In the latter case the elementary excitation particles of vibrations
of crystal lattice or quanta of the sound field are named phonons \cite{Einstein1907,Debye1912}.
Both of these quantum fields are bosonic ones \cite{Feynman1972,FetterWalecka1971}.
The harmonic potential energy operator is only an approximation for the real anharmonic
potential energy operator of mutual interaction between atoms (ions, nuclei) in real crystals.
Therefore the motivation of the present paper is an attempt for a more realistic description
of quantum anharmonical systems.

\section{Quantum harmonic oscillator in $D=3$ spatial dimensions}

{\sl Firstly:} Quantum harmonic oscillator in $D=3$ spatial dimensions 
is a solvable analytically model.
In order to make the deliberations easier we reduce our present interest 
to the first quantization case.
Therefore the relevant Hilbert space ${\cal V}_{3}$ 
of the complex-valued wave functions that are modulus square integrable
on the set ${\bf R}^{3}$ of $(D=3)$-tuples (triples) ${\bf x}$ of real numbers,
is isomorphic 
to a $(D=3)$-dimensional tensor (Cartesian) product
of the Hilbert spaces ${\cal V}_1$ of the complex-valued wave functions that are modulus square integrable
on the set ${\bf R}$:
\begin{equation}
{\cal V}_{3} = L^2({\bf R}^{3}, {\bf C}, {\rm d}{\bf x})\equiv \bigotimes_{j=1}^{3} {\cal V}_1,
\label{Hilbert-space-L2-D=3-tensor-product}
\end{equation}
where
\begin{equation}
{\cal V}_1 = L^2({\bf R}, {\bf C}, {\rm d}x_j),
\label{Hilbert-space-L2-D=1}
\end{equation}
whereas ${\bf R}$ is the set of the real numbers,
and ${\bf C}$ is the set of the complex numbers.
The ${\bf r}$th harmonic oscillator's eigenfunction $\Psi_{\bf r}$ 
in $D=3$ dimensions is, neither symmetrized nor antisymmetrized, 
tensor product of the eigenfunctions in one dimension:
\begin{equation}
\Psi_{\bf r}({\bf x}) = \prod_{j=1}^{3} \Psi_{r_{j}}(x_j), {\bf x}=(x_1,...,x_{D=3}) \in {\bf R}^{3}, 
{\bf r}=(r_1,...,r_{D=3}) \in {\bf N}^{3},
\label{eigenfunction-tensor-product-definition-D=3}
\end{equation}
where we used boldface font for the $D=3$-tuples ${\bf x},$ and ${\bf r}.$
The Hermite's functions $\Psi_{r_j}$ 
(Fock's functions, eigenfunctions of the dimensionless 
Hamiltonian operator $\hat{{\cal H}}_{(j)}$ 
of the quantum harmonic oscillator in $D=1$ spatial dimension) read:
\begin{equation}
\Psi_{r_j}(x_j) = N_{r_j} H_{r_j}(x_j) \exp(-\frac{1}{2}x_j^2), 
N_{r_j} = [ \sqrt{\pi} r_j! 2^{r_j} ]^{-\frac{1}{2}}, r_j \in {\bf N},
\label{Hermite-function-definition}
\end{equation}
where ${\bf N}$ is a set of natural numbers including zero,
whereas:
\begin{equation}
H_{r_j}(x_j) = (-1)^{r_j} \exp(x_j^2) \frac{{\rm d}^{r_j}}{{\rm d}x_j^{r_j}} \exp(-x_j^2), 
\label{Hermite-polynomial-definition}
\end{equation}
is $r_j$th Hermite's polynomial \cite{Davydov1963}.
It follows that:
\begin{equation}
\Psi_{\bf r}({\bf x}) = N_{\bf r} H_{\bf r}({\bf x}) \exp(-\frac{1}{2}{\bf x}^2), 
N_{\bf r} = \prod_{j=1}^{3} N_{r_{j}},  
H_{\bf r}({\bf x})= \prod_{j=1}^{3} H_{r_{j}}(x_j),
\label{eigenfunction-Hermite-function-definition-D=3}
\end{equation}
whereas 
\begin{equation}
{\bf x}^2={\bf x} \cdot {\bf x}= \sum_{j=1}^{3}(x_j)^2.
\label{D=3-tuple-scalar-product}
\end{equation}
One can also draw a conclusion that the Hilbert space:
\begin{equation}
{\cal V}_{3} = L^2({\bf R}^{3}, {\bf C}, {\rm d}x), 
\label{Hilbert-space-L2-D=3}
\end{equation}
is composed of the complex-valued wave functions $\Psi$ that are modulus square integrable
on the set ${\bf R}^{3}$.
The Hilbert space ${\cal V}_{3}$ is separable space,
and its orthonormal basis ${\cal B}_{3}$ is a set of Hermite's functions $\Psi_{\bf r}$ in $D=3$ dimensions 
(Fock's functions in $D=3$ dimensions).
The {\sl dimensionless (nondimensional)} 
quantum Hamiltonian operator $\hat{{\cal H}}_{3}$ of the quantum harmonic oscillator in $D=3$ dimensions 
is a sum of quantum Hamiltonian operators $\hat{{\cal H}}_{(j)}$ 
of the quantum harmonic oscillators in one dimension:
\begin{equation}
\hat{{\cal H}}_{3} = \sum_{j=1}^{3} \hat{{\cal H}}_{(j)} 
= \sum_{j=1}^{3} (\hat{p}_j^2+ \hat{x}_{j}^2) = \hat{{\bf p}}^2 + \hat{{\bf x}}^2. 
\label{H-operator-harmonic-D=3-definition}
\end{equation}  
The {\sl dimensionless (nondimensional)} 
quantum Hamiltonian operator $\hat{{\cal H}}_{(j)}$ of the quantum harmonic operator
in one spatial dimension is defined as follows:
\begin{equation}
\hat{{\cal H}}_{(j)} = \hat{p}_j^2 + \hat{x}_j^2. 
\label{H_j-operator-harmonic-1D-definition}
\end{equation}  
The quantum Hamiltonian operator $\hat{{\cal H}}_{3}$ is diagonal in Fock's basis of its
eigenfunctions $\Psi_{\bf r}$:
\begin{equation}
\hat{{\cal H}}_{3} \Psi_{\bf r} = 
\epsilon_{\bf r} \Psi_{{\bf r}}, 
\label{H-operator-harmonic-D=3-Hermite-function-action}
\end{equation}
and its matrix element $({\cal H}_{3})_{{\bf l}, {\bf r}}$ is equal:
\begin{equation}
({\cal H}_{3})_{{\bf l}, {\bf r}} = 
\epsilon_{\bf r} \delta_{{\bf l},{\bf r}}, 
\label{H-operator-harmonic-D=3-Hermite-function-matrix-element}
\end{equation} 
whereas
\begin{equation}
\epsilon_{\bf r}  
= \sum_{j=1}^{3} \epsilon_{r_{j}} = \sum_{j=1}^{3} (2 r_{j}+1), 
\label{H-operator-harmonic-D=3-Hermite-function-eigenvalue}
\end{equation} 
is the ${\bf r}$th eigenenergy of $\hat{{\cal H}}_{3}$ 
and where
\begin{equation}
\delta_{{\bf l},{\bf r}} = \prod_{j=1}^{3} \delta_{l_j,r_j}, 
\label{Kronecker-delta-definition-D=3}
\end{equation}
is discrete Kronecker's delta in $D=3$ dimensions 
(it is not continuous Dirac's delta in $D=3$ dimensions).
The eigenenergies $\epsilon_{\bf r}$ are simply the sums of all odd natural numbers, 
and the quantum Hamiltonian $\hat{{\cal H}}_{3}$ 
is (direct) sum of diagonal operators in one dimension, 
and its matrix representation is direct sum of diagonal $\infty \times \infty$ matrices
(it is poly-index matrix).

{\sl Secondly,} let us perform very difficult task 
consisting of calculating all the flip-flop transition amplitudes (hopping amplitudes) 
from the quantum state $\chi_{\bf r}^{\bf s} = \hat{{\bf x}}^{\bf s} \Psi_{\bf r}$ 
to the quantum state $\Psi_{\bf l}$ (${\bf s} \geq {\bf 0}$):
\begin{equation}
(m_{\bf s})_{{\bf l}, {\bf r}}=({\bf x}^{\bf s})_{{\bf l}, {\bf r}} 
= \langle \Psi_{\bf l} | \hat{{\bf x}}^{\bf s} \Psi_{\bf r} \rangle_{{\cal V}_{3}} 
= \prod_{j=1}^{3}\int_{-\infty}^{\infty} \Psi_{l_{j}}^{\star}(x_j) x_j^{s_{j}}
\Psi_{r_{j}}(x_j){\rm d}x_j.
\label{x-operator-sth-transition-amplitude-D=3}
\end{equation} 
The flip-flop transition amplitude $(m_{\bf s})_{{\bf l}, {\bf r}}$ is connected with
the processes of emissions and/or absorptions of ${\bf s}$ phonons in $D=3$ spatial dimensions,
because:
\begin{equation}
\hat{{\bf x}}^{\bf s} = \prod_{j=1}^{3} (x_j^{s_{j}})
=\prod_{j=1}^{3} [\sqrt{2}^{s_j} (\hat{a}_j + \hat{a}_j^{+})^{s_j})]
=[\sqrt{2} (\hat{{\bf a}} + \hat{{\bf a}}^{+})]^{\bf s}, 
\label{x-operator-annihilation-creation-operators-relation-D=3}
\end{equation} 
where $\hat{{\bf a}}=(\hat{a}_1,...,\hat{a}_{D=3}), 
\hat{{\bf a}}^{+}=(\hat{a}_1^{+},...,\hat{a}_{D=3}^{+}),$
are the bosonic multiphonon ($(D=3)$-phonon) 
annihilation and creation operators in $D=3$ spatial dimensions, respectively, and:
\begin{equation}
\hat{a}_j \Psi_{\bf r} = \sqrt{r_j} \Psi_{(r_1,...,r_j-1,...,r_{D=3})},
\hat{a}_j^{+} \Psi_{\bf r} = \sqrt{r_j+1} \Psi_{(r_1,...,r_j+1,...,r_{D=3})}. 
\label{x-operator-annihilation-creation-operators-definition-D=3}
\end{equation} 
It can be easily proven that:
\begin{equation}
({\bf x}^{\bf s})_{{\bf l}, {\bf r}} 
= \prod_{j=1}^{3} (x_j^{s_{j}})_{l_j, r_j} . 
\label{x-operator-sth-power-definition-D=3}
\end{equation} 
One can calculate the lower transition amplitudes manually, {\sl e. g.}, 
using recurrence relations, matrix algebra, {\sl etc.},
but it is tedious (even for $3 \leq s_j \leq 6$). 
The exact formula for {\sl all} the transition amplitudes reads:
\begin{equation}
(m_{\bf s})_{{\bf l}, {\bf r}}=\prod_{j=1}^{3} (m_{s_{j}})_{l_j, r_j}.
\label{x-operator-sth-transition-amplitude-matrix-element-D=3} 
\end{equation}
One can calculate the lower transition amplitudes manually $(m_{s_{j}})_{l_j, r_j}$, {\sl e. g.}, 
using recurrence relations, matrix algebra, {\sl etc.},
but it is tedious (even for $3 \leq s_j \leq 6$). 
If one wants to calculate {\sl all} the transition amplitudes then he must return 
to the beautiful XIX century mathematics methods
and after some reasoning he obtains the exact formula \cite{Duras1996}:
\begin{eqnarray}
& & (m_{s_j})_{l_j,r_j}= 
\nonumber \label{x_j-operator-s_jth-transition-amplitude-matrix-element-1} \\ 
& & = [1 - (-1)^{s_j+l_j+r_j}] \sum_{\lambda_j=0}^{[l_j/2]} \sum_{\rho_j=0}^{[r_j/2]} 
[ (-1)^{\lambda_j+\rho_j} 
\frac{\sqrt{l_j!}}{\lambda_j! (l_j-2\lambda_j)!} 
\cdot \frac{\sqrt{r_j!}}{\rho_j! (r_j-2\rho_j)!} \cdot 
\nonumber \label{x_j-operator-s_jth-transition-amplitude-matrix-element-2} \\
& & \cdot 2^{\frac{l_j}{2}+\frac{r_j}{2}-2\lambda_j-2\rho_j-1}
\cdot \Gamma(\frac{s_j+l_j+r_j-2\lambda_j-2\rho_j+1}{2}) ], 
\label{x_j-operator-s_jth-transition-amplitude-matrix-element-3} 
\end{eqnarray}
where $[\cdot]$ is entier (step) function, $\Gamma$ is Euler's gamma function
(compare our result Eq. (\ref{x_j-operator-s_jth-transition-amplitude-matrix-element-3}) 
with the formulae in \cite{Graffi1973,Balsa1983}).

\section{Quantum anharmonic oscillator in $D=3$ spatial dimensions}

{\sl Thirdly,} we are ready to investigate the quantum anharmonic oscillator 
in $D=3$ spatial dimensions.
Its {\sl dimensionless} Hamiltonian operator 
$\hat{{\cal H}}_{3, {\rm anharm}}^{{\bf S}}$ reads:
\begin{equation}
\hat{{\cal H}}_{3, {\rm anharm}}^{{\bf S}} 
=\hat{{\cal H}}_{3}+ \sum_{{\bf s}={\bf 0}}^{{\bf S}} a_{\bf s} \hat{{\bf x}}^{\bf s}
=\hat{{\cal H}}_{3}+ \sum_{(s_1,...,s_{D=3})=(0,...,0)}^{(S_1,...,S_{D=3})} [a_{(s_1,..,s_{D=3})} \cdot 
\prod_{j=1}^{3}(\hat{x}_j)^{s_j}], 
\label{H-operator-anharmonic-D=3-S}
\end{equation}
where ${\bf S}$ is a $D=3$-tuple of degrees of the anharmonicity of the oscillator, 
and the prefactors $a_{\bf s}$ are the strengths of anharmonicity.
The matrix elements of the anharmonic Hamiltonian operator are:
\begin{equation}
({\cal H}_{3, {\rm anharm}}^{{\bf S}})_{{\bf l}, {\bf r}} 
=\epsilon_{\bf r} \delta_{{\bf l}, {\bf r}} 
+ \sum_{{\bf s}={\bf 0}}^{{\bf S}} a_{\bf s} ({\bf{x}}^{\bf s})_{{\bf l}, {\bf r}}
=\epsilon_{\bf r} \delta_{{\bf l}, {\bf r}} 
+ \sum_{{\bf s}={\bf 0}}^{{\bf S}} a_{\bf s} ({\bf{m}}_{\bf s})_{{\bf l}, {\bf r}}, 
\label{H-operator-anharmonic-D=3-S-matrix-element}
\end{equation}
where the representation of the $D=3$-dimensional quantum anharmonic oscillator 
in the quantum harmonic oscillator basis ${\cal B}_{3}$
is mathematically correct, because the basis ${\cal B}_{3}$ is a complete set, 
and the Hilbert space of the eigenfunctions
of the anharmonic oscillator is isomorphic 
to the Hilbert space ${\cal V}_{3}$ for the harmonic oscillator, 
provided that
the total potential energy ${\cal U}_{3, {\rm total}}^{{\bf S}}$ 
of the quantum anharmonic oscillator in $D=3$ dimensions: 
\begin{equation}
{\cal U}_{3, {\rm total}}^{{\bf S}}({\bf x}) 
= {\bf x}^2+ {\cal U}_{3, {\rm anharm}}^{{\bf S}}({\bf x}), 
\label{U-total-potential-energy-D=3-S}
\end{equation}
is bounded from below (there are no scattering eigenstates in $D=3$ dimensions),
where the anharmonic potential energy ${\cal U}_{3, {\rm anharm}}^{{\bf S}}$ is:
\begin{equation}
{\cal U}_{3, {\rm anharm}}^{{\bf S}}({\bf x}) = \sum_{{\bf s}={\bf 0}}^{{\bf S}} a_{\bf s} {\bf x}^{\bf s}
=\sum_{(s_1,...,s_{D=3})=(0,...,0)}^{(S_1,...,S_{D=3})} [a_{(s_1,..,s_{D=3})} \cdot 
\prod_{j=1}^{3}(x_j)^{s_j}]. 
\label{U-anharmonic-potential-D=3-S}
\end{equation}
It suffices that the $D=3$-tuple of degrees of the anharmonicity of the oscillator 
${\bf S}=2 {\bf S'}$ is composed of even numbers 
and that the strength of anharmonicity $a_{\bf S}$ is strictly positive: 
$a_{\bf S} > 0$, so that ${\cal U}_{3, {\rm total}}^{{\bf S}}({\bf x}) \rightarrow \infty$
for $|{\bf x}| \rightarrow \infty$.

{\sl Fourthly,} we repeat the ``Bohigas conjecture'' 
that the fluctuations of the spectra of the quantum systems
that correspond to the chaotic systems generally obey 
the spectra of the Gaussian random matrix ensembles.
The quantum integrable systems correspond to the classical integrable systems 
in the semiclassical limit
\cite{Bohigas1984,OzoriodeAlmeida1988,Haake1990,Guhr1998,Mehta19900,Reichl1992,Bohigas1991,Porter1965,Brody1981,Beenakker1997,Ginibre1965,Mehta19901}. 
We emphasize that the ``Bohigas conjecture'' also holds 
for the quantum oscillators in $D=3$ dimensions. 
Having conducted many numerical experiments with different quantum anharmonic oscillators 
(up to the sextic $(D=3)$-dimensional quantum anharmonic oscillators: $S_j=6$) 
we draw conclusion that some of them behave like quantum
integrable systems, the eigenenergies tend to cluster, 
the histograms of nearest neighbour spacing are closer to
the $P_0$ distribution resulting from the Poisson ensemble,
whereas other ones look like quantum chaotic systems, 
their eigenenergies are subject to repulsion,
the histograms of NNS are closer 
to the distributions $P_1, P_2, P_4,$ derived from the Gaussian
Random Matrix ensembles \cite{Duras1996}.




\begin{thebibliography}{1}

\bibitem{Einstein1905} 
A. Einstein, 
{\em Annalen der Physik (Leipzig)} {\bf 17}, p. 132 (1905). 
\bibitem{Lewis1926} 
G. N. Lewis, 
Nature {\bf 118}, p. 874 (1926). 
\bibitem{Einstein1907} 
A. Einstein, 
{\em Annalen der Physik (Leipzig)} {\bf 22}, p. 180 (1907). 
\bibitem{Debye1912} 
P. Debye, 
{\em Annalen der Physik (Leipzig)} {\bf 39}, p. 789 (1912). 
\bibitem{Feynman1972} 
R. P. Feynman, 
{\em  Statistical Mechanics: A Set of Lectures}, 
W. A. Benjamin, Reading, Massachusetts, 1972.
\bibitem{FetterWalecka1971} 
A. L. Fetter, J. D. Walecka, 
{\em Quantum theory of Many-Particle Systems}, 
McGraw-Hill Book Company, San Francisco, 1971.
\bibitem{Davydov1963}
A. S. Davydov, 
{\em Quantum Mechanics}, 
GIFML Editors, Moscow, 1963.
\bibitem{Graffi1973} 
S. Graffi, V. Grecchi, 
{\em Phys. Rev.} D {\bf 8}, p. 3487 (1973). 
\bibitem{Balsa1983} 
R. Balsa, M. Plo, J. G. Esteve, A. F. Pacheco, 
{\em Phys. Rev.} D {\bf 28}, p. 1945 (1983). 
\bibitem{Bohigas1984} 
O. Bohigas, M. J. Giannoni, C. Schmidt, 
{\em Phys. Rev. Lett.} {\bf 52}, p. 1 (1984). 
\bibitem{OzoriodeAlmeida1988} 
A. M. Ozorio de Almeida, 
{\em Hamiltonian systems: chaos and quantization}, 
Cambridge University Press, Cambridge, 1988.
\bibitem{Haake1990}
 F. Haake, 
{\em Quantum Signatures of Chaos},
Springer-Verlag, Berlin, Heidelberg, New York, 1990, 
Chapters 1, 3, 4, 8, pp. ~1--11, 33--77, 202--213.
\bibitem{Guhr1998} 
T. Guhr, A. M\"uller-Groeling, H. A. Weidenm\"uller, 
{\em Phys. Rep.} {\bf 299}, p. 189 (1998).  
\bibitem{Mehta19900}
M. L. Mehta, 
{\em Random matrices}, 
Academic Press, Boston, 1990, Chapters 1, 2, 9, pp. ~1--54, 182--193.
\bibitem{Reichl1992}
L. E. Reichl, 
{\em The Transition to Chaos In Conservative Classical 
 Systems: Quantum Manifestations}, Springer-Verlag, New York, 1992, 
 Chapter 6, pp. ~248--286.
\bibitem{Bohigas1991}
 O. Bohigas, in {\em Proceedings of the Les Houches Summer School on Chaos and
 Quantum Physics, Session LII, 1 - 31 August 1989 
 [LES HOUCHES \'ECOLE D'\'ET\'E DE PHYSIQUE TH\'EORIQUE, 
 SESSION LII, 1 - 31 AO\^UT 1989]} 
 edited by M. - J. Giannoni, A. Voros, J. Zinn-Justin 
 North-Holland, Amsterdam, 1991, pp. ~87--199.
\bibitem{Porter1965}
C. E. Porter, 
{\em Statistical Theories of Spectra: Fluctuations}, 
Academic Press, New York, 1965.
\bibitem{Brody1981}
T. A. Brody, J. Flores, J. B. French, P. A. Mello, A. Pandey, S. S. M. Wong, 
{\em Rev. Mod. Phys.} {\bf 53}, p. 385 (1981).
\bibitem{Beenakker1997}
C. W. J. Beenakker, 
{\em Rev. Mod. Phys.} {\bf 69}, p. 731 (1997).  
\bibitem{Ginibre1965} 
J. Ginibre,  
{\em J. Math. Phys.} {\bf 6}, p. 440 (1965). 
\bibitem{Mehta19901}
M. L. Mehta, 
{\em Random matrices}, 
Academic Press, Boston, 1990, Chapter 15, pp. ~294--310.
\bibitem{Duras1996}
M. M. Duras, unpublished. 

\end{thebibliography}
\end{document}